\documentclass{article}

\usepackage{arxiv}
\usepackage[utf8]{inputenc}
\usepackage{cite}
\usepackage[right]{lineno}
\usepackage{microtype}
\DisableLigatures[f]{encoding = *, family = * }
\raggedright
\usepackage{changepage}
\usepackage[aboveskip=1pt,labelfont=bf,labelsep=period,singlelinecheck=off]{caption}
\makeatletter
\renewcommand{\@biblabel}[1]{\quad#1.}
\makeatother
\usepackage{lastpage,fancyhdr,graphicx}
\usepackage{epstopdf}
\fancyheadoffset[L]{2.25in}
\fancyfootoffset[L]{2.25in}
\usepackage{color}
\definecolor{Gray}{gray}{.25}
\usepackage{graphicx}
\usepackage{sidecap}
\usepackage{wrapfig}
\usepackage[pscoord]{eso-pic}
\usepackage[fulladjust]{marginnote}
\reversemarginpar
\usepackage{chemstyle}

\title{Molecular Mechanism of Gas Solubility in Liquid: Constant Chemical Potential Molecular Dynamics Simulations}

\author{
  Narjes Ansari\\
  Department of Chemistry and Applied Biosciences, ETH Z\"{u}rich\\
  Facolt\`{a} di informatica, Istituto di Scienze Computazionali\\
  Università della Svizzera Italiana, CH-6900 Lugano, Switzerland\\
   \And
 Tarak Karmakar \\
 Department of Chemistry and Applied Biosciences, ETH Z\"{u}rich\\
  Facolt\`{a} di informatica, Istituto di Scienze Computazionali\\
  Università della Svizzera Italiana, CH-6900 Lugano, Switzerland\\
  \And
  Michele Parrinello\\
   Department of Chemistry and Applied Biosciences, ETH Z\"{u}rich\\
  Facolt\`{a} di informatica, Istituto di Scienze Computazionali\\
  Università della Svizzera Italiana, CH-6900 Lugano, Switzerland\\
  Italian Institute of Technology, Via Morego 30, 16163, Genova, Italy\\
    \texttt{parrinello@phys.chem.ethz.ch} \\
}

\begin{document}
\maketitle

\begin{abstract}
%\lipsum[1]
{Accurate prediction of a gas solubility in a liquid is crucial in many areas of chemistry, and a  detailed understanding of the molecular mechanism of the gas solvation continues to be an active area of research. Here, we extend the idea of constant chemical potential molecular dynamics (C$\mu$MD) approach to the calculation of the gas solubility in the liquid under constant gas chemical potential conditions. As a representative example, we utilize this method to calculate the isothermal solubility of carbon dioxide in water. Additionally, we provide microscopic insight into the mechanism of solvation that preferentially occurs in areas of the surface where the hydrogen network is broken.}
\end{abstract}

% keywords can be removed

\newpage
\section{Introduction}
Solubility measures the ability of a substance in its solid, liquid, or gaseous state to dissolve into a solvent. Accurate prediction of solubility and a detailed understanding of the molecular mechanism of solute penetration across the interfaces is crucial in many areas of chemistry, geochemistry, inorganic, physical, organic, and biochemistry \cite{tomkins2008applications,constable2007perspective,chen2018prediction}. For example, drug solubility is a crucial parameter in the development of active pharmaceuticals \cite{savjani2012drug,liu2018water}. In the petroleum industry, design of purification process and chemical separation depends on relative solubilities~\cite{takahashi2007tunable,chen2005calcium}. Solubility in water is of special interest since most fundamental biological, pharmacological, industrial, and environmental functions take place in aqueous solvent. \cite{amidon1995theoretical,jorgensen2002prediction,jorgensen2000prediction,jain2001estimation,ishikawa2011improvement}.

Much effort has been devoted to developing successful computational tools for solubility prediction~\cite{delaney2005predicting,hossain2019molecular,li2017computational,boothroyd2018solubility}. Free energy methods such as free energy perturbation~\cite{zwanzig1954high} and thermodynamic integration~\cite{kirkwood1935statistical} are widely used~\cite{duarte2017approaches,ferrario2002solubility,sanz2007solubility}. In these methods, the solvation free energy is computed from the free energy change associated with the transfer of a solute molecule from an ideal gas to the solvent. However, the main drawback of these methods is the neglect of the solute-solute interactions both in the gas and the solvent phases.

An alternative and more realistic way of approaching this problem is to use a standard vapor-liquid equilibrium (VLE) molecular dynamics (MD) simulation. In this method, the solute molecules are allowed to diffuse into the solvent, and the solubility is obtained from the equilibrium solute density inside the liquid. However, during the solvation process, the density of the solute in the vapor phase is depleted, causing a decrease in its chemical potential. This effect can eventually lead to an underestimation of the solubility. To avoid such an artifact, one could keep the pressure constant by using an NPT simulation. Unfortunately, for the system studied here, namely the solvation of CO$_{2}$ in water, one peculiar finite-size effect takes place. That is, when the number of gas molecules is small the system makes a transition to a layered structure in which water and liquid CO$_{2}$ alternate (see Section I of the SI). When the number of gas molecules is increased this effect disappears but this forces one to use a large system. Not only that, but given the high gas compressibility, one has to deal with large volume fluctuations. 

A possible way of avoiding this artifact is to use a grand-canonical ($\mu$VT) ensemble. In this approach, the simulation box is coupled to a reservoir that supplies the solute molecules to the system, and thus, one is able to control the solute density via a Monte Carlo procedure. Here we explore an alternative to this approach using C$\mu$MD~\cite{perego2015molecular}. This is a method recently developed in our group in order to perform constant chemical potential simulations. The method has been successfully tested in the study of crystallization from solution~\cite{karmakar2018,karmakar2019,bjelobrk2019,han2019solvent}, and gas permeation in organic membranes~\cite{ozcan2017concentration,ozcan2020modelling}. Here, we adopt this method to simulate the solvation of the gas in the liquid under constant chemical potential conditions.

To test the effectiveness of our C$\mu$MD method, we applied it to determine the isothermal solubility of CO$_{2}$ in water in a range of pressures and temperatures. The availability of a substantial amount of experimental and theoretical result makes this system a suitable test of the method~\cite{takenouchi1964binary,blencoe2004co2,iacono2012new,sujith2020adsorption,mao2013improved,spycher2003co2,duan2003Thermo,liu2011monte,moultos2014atomistic}. Moreover, the CO$_{2}$ - H$_{2}$O system is one of the fundamental geochemical binary systems, and knowledge of the gas solvation process is crucial for furthering our fundamental understanding and to advance technological and environmental applications~\cite{metz2005carbon,nordbotten2011geological}. 

An additional benefit of a two-fluid simulation is that one can obtain information on how the solute molecule dissolves in the liquid. This is a subject of great relevance to which much effort has been devoted~\cite{ishiyama2014theoretical, godawat2009characterizing,saykally2013air, vacha2012charge,zhang2011analysis,sakamaki2011thermodynamic,pezzotti20172d,godawat2009characterizing,wang2018ion,sujith2020adsorption}. In this work, by analyzing the local curvature of water interface, we find that the distribution of water molecules is not uniform and depends on surface curvature. A detailed analysis reveals that
near this instantaneous interface, there are transient empty patches. These patches are mostly localized at the crest of the interface and provide an entry point for the gas molecules into the liquid.

\section{Computational Details}
In this section, we first briefly describe the C$\mu$MD method followed by the details of the thermodynamic conditions and MD simulations.\\

\subsection{C$\mu$MD Method} 
\begin{figure}[!ht]
  \begin{center}
    \includegraphics[width=0.7\textwidth]{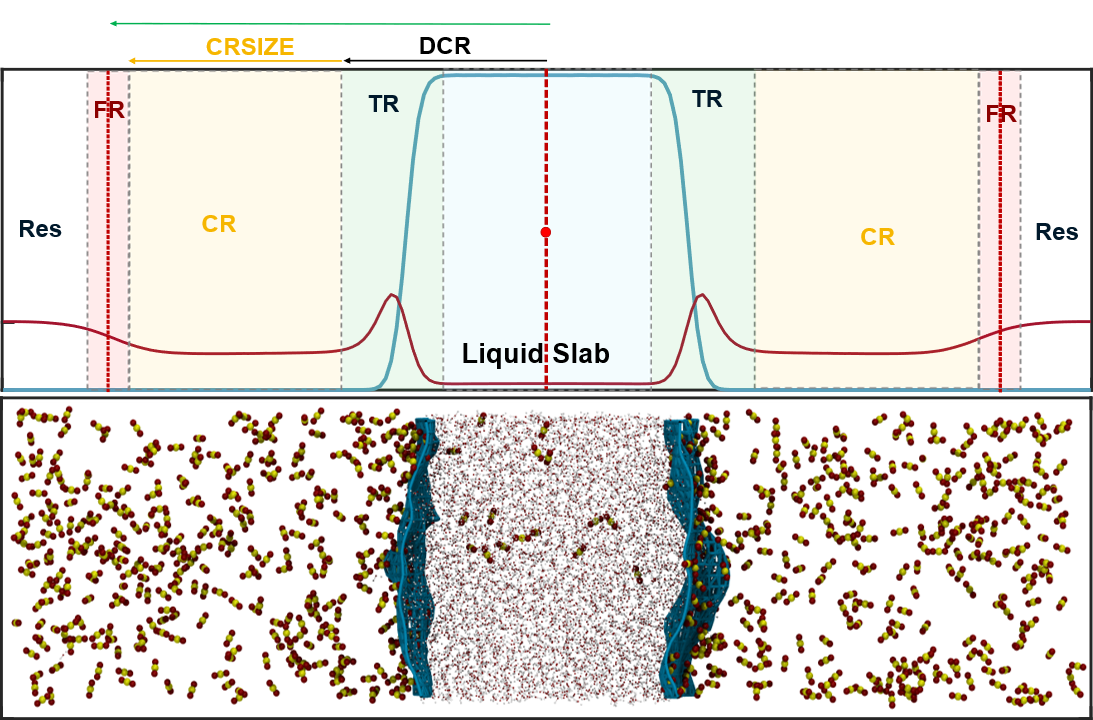}
  \end{center}
  \caption{a) Schematic representation of the C$\mu$MD simulation setup, (b) Snapshot of the C$\mu$MD simulation box of CO$_{2}$ - H$_{2}$O binary system. The simulation box is symmetric with the liquid in the middle.  The red point at the middle of the box is the liquid slab center of mass and all partitioning of the gas chamber is respected to this point. Two Willard-Chandler interfaces are shown in cyan. }
  \label{CmuMD_shem}
\end{figure}

In this work, we use an orthorhombic simulation box and apply periodic boundary conditions in all directions. A liquid slab sits in the middle of the simulation box (Fig.~\ref{CmuMD_shem}(a) and (b)). The two sides of the liquid slab are filled with the gas molecules. The gas chamber is divided into different regions (see Fig.~\ref{CmuMD_shem}(a)). On each side of the liquid slab we define control regions (CR, shown in orange in Fig.~\ref{CmuMD_shem}(a)), where the number density of the gas is controlled by an external force (see below). The region between the liquid slab and the CR is called the transition region (TR). Beyond the force region (FR) lies a molecular reservoir (Res) that maintains the CR density constant by supplying the extra molecules.

The first step in the C$\mu$MD algorithm is to calculate the solute number density $c$ in the CR as,
\begin{equation}
c = \frac{1}{V^{CR}} \sum^{N}_{j=1} f(z_j),
\label{eq:nCR}
\end{equation}
where $V^{CR}$ is the CR volume and $z_j$ the distance of a j-th particle from the box center. The $f(z_j)$ is a continuous and differentiable switching function that counts the number of atoms in the CR. In our case, this function is defined as a product of two Fermi switching functions, as:
\begin{equation}
%\begin{split}
 f(z_j) =\frac{1}{1+e^{-(z_j-z_{in})/\alpha}}.\frac{1}{1+e^{(z_j-z_{out})/\alpha}}, 
%\end{split}
\label{eq:fermi}
\end{equation}
where, $z_{in}$ and $z_{out}$ are the inner and outer CR boundaries and $\alpha$ is a parameter that controls the switching functions steepness (Fig. ~\ref{CmuMD_shem}(a)). The function $f(z_j)$ has a value of 1 when the solute is inside the CR and 0 when outside. From now on, we drop the atomic index for simplicity.

Now, we apply a force on the instantaneous solute density ($c$ in Eq.~\ref{eq:nCR}) such that it remains to the target value ({\it $c_0$}), 
\begin{equation}
F(r) = \kappa(c-c_{0})G(z)
\label{eq:F}
\end{equation}
where $\kappa$ is the force constant. The $G(z)$ in Eq. (3) is a bell-shaped function that localizes the {\it $F(z)$} at $z_F$. 
This function ($G(z)$) is defined as, \begin{equation}
	G = \frac{1}{2\sigma}\left [ \frac{1}{1+\cosh \left(\frac{z-z_F}{\sigma}\right)} \right ]
\end{equation}
where $\sigma$ is a broadening parameter. 

The size of each region, the number density of the solute in the reservoir and the force constant are among the parameters that need to be chosen. The regions (TR, CR, FR, and the Res) should be defined in such a way that they do not overlap. Details of the C$\mu$MD parameters used in our simulations are provided in Section II of the SI.\\

\noindent
\subsection{Potential Model and Thermodynamic Conditions}

Besides using an appropriate computational tool, the accuracy of the result depends on the quality of the model interaction potential~\cite{vorholz2000vapor,huang2009henry, vorholz2004molecular,liu2011monte, liu2013simulations}. An investigation of the performance of different models has been carried out by Liu et al. using histogram-reweighting Grand Canonical Monte Carlo (GCMC)~\cite{liu2011monte}.
Their results show that none of the potentials tested was able to reproduce the experimental results over the range of pressure and temperature investigated. In this paper, we use for water the SPC~\cite{harris1995carbon} model and EPM2~\cite{harris1995carbon} for CO$_{2}$. The Lorentz–Berthelot combination rules~\cite{allen2017computer} were applied to determine the CO$_{2}$ - H$_{2}$O interaction parameters. With this choice, a good agreement is obtained for T=423 K~\cite{liu2011monte}. Guided by the Liu et al. experience we chose to investigate at this temperature and the pressure P=100 K (system \emph{I}), where also experimental data is available, and the system is supercritical. We also moved close to the experimental critical point (T=304.13 K, P=73.8 bar) performing a second simulation at T=323 K, P=50 bar (system \emph{II}), where CO$_{2}$ is in the vapor phase.\\

\noindent
\subsection{Simulation Protocol}
To prepare the initial configurations, we follow three steps: 1) we simulate the bulk liquid in a cubic box of 4055 water molecules at the chosen thermodynamic conditions, 2) if $a$ is the equilibrium value of the initial cubic box, the simulation box is then adiabatically elongated along the z-axis until the final orthorhombic box has dimension $a$ $\times$ $a$ $\times$ 4$a$ (see Table S1), 3) the space left empty by water is filled by CO$_{2}$ molecules. The number of the gas molecules is fixed to the desired value.

Fig.~\ref{CmuMD_shem}(b) shows a snapshot of the simulation box. The first two steps are identical in the C$\mu$MD and NVT-MD simulations, while in the third step the number of CO$_{2}$ molecules are different. In the NVT-MD simulations, the number of CO$_{2}$ molecules in the vapor phase is chosen so as to reproduce the experimental CO$_{2}$ gas density at the desired thermodynamic conditions. In the C$\mu$MD simulations, the number of CO$_{2}$ in the CR and TR is kept close to that used in the NVT-MD simulations, while in the Res this number is $\sim$1.3 times higher (see Table S1). As mentioned earlier, the C$\mu$MD simulations involve several parameters such as the size of TR, CR, FR, Res, and the value of $\kappa$. The C$\mu$MD parameters are chosen after performing a small number of simulations (see Table S2).  
All simulations were carried out using the GROMACS-2018.3 software~\cite{abraham2015}. The C$\mu$MD method is implemented in a private version of the PLUMED2.6 code~\cite{tribello2014plumed}. The Ewald\cite{Essmann_JCP_1995_PME} summation was used for the long-range electrostatic interactions, and a cutoff of 10 \AA \ was used for both the Coulomb and the van der Waals interactions. A time step of 2 fs was used in all simulations. A pure bulk water system was equilibrated at each pressure for 10 ns in the NPT ensemble using the Parrinello-Rahman barostat~\cite{ParinelloRahman_JAP_1981_PR-barostat} and the stochastic velocity rescaling thermostat~\cite{bussi2007canonical}. For both thermostat and barostat, the coupling constant was 1 ps. Then, the vapor-liquid systems were equilibrated for 100 ps, and statistics was accumulated for 100 ns. In order to understand the properties of the interface, a 100 ns long simulation of the vapor-liquid water system was carried out. 

For the solubility calculation, both water and CO$_{2}$ equilibrated densities were measured in the region of $\sim$3 nm width around the center of water density profile. The analysis was performed using 10$^{4}$ configurations extracted at 10 ps intervals from the equilibrium segment of the trajectory. The Visual Molecular Dynamics (VMD) software~\cite{humphrey1996vmd} was used to visualize the trajectories and produce some of the figures.

\begin{figure}[!ht]
\includegraphics[width=0.6\textwidth]{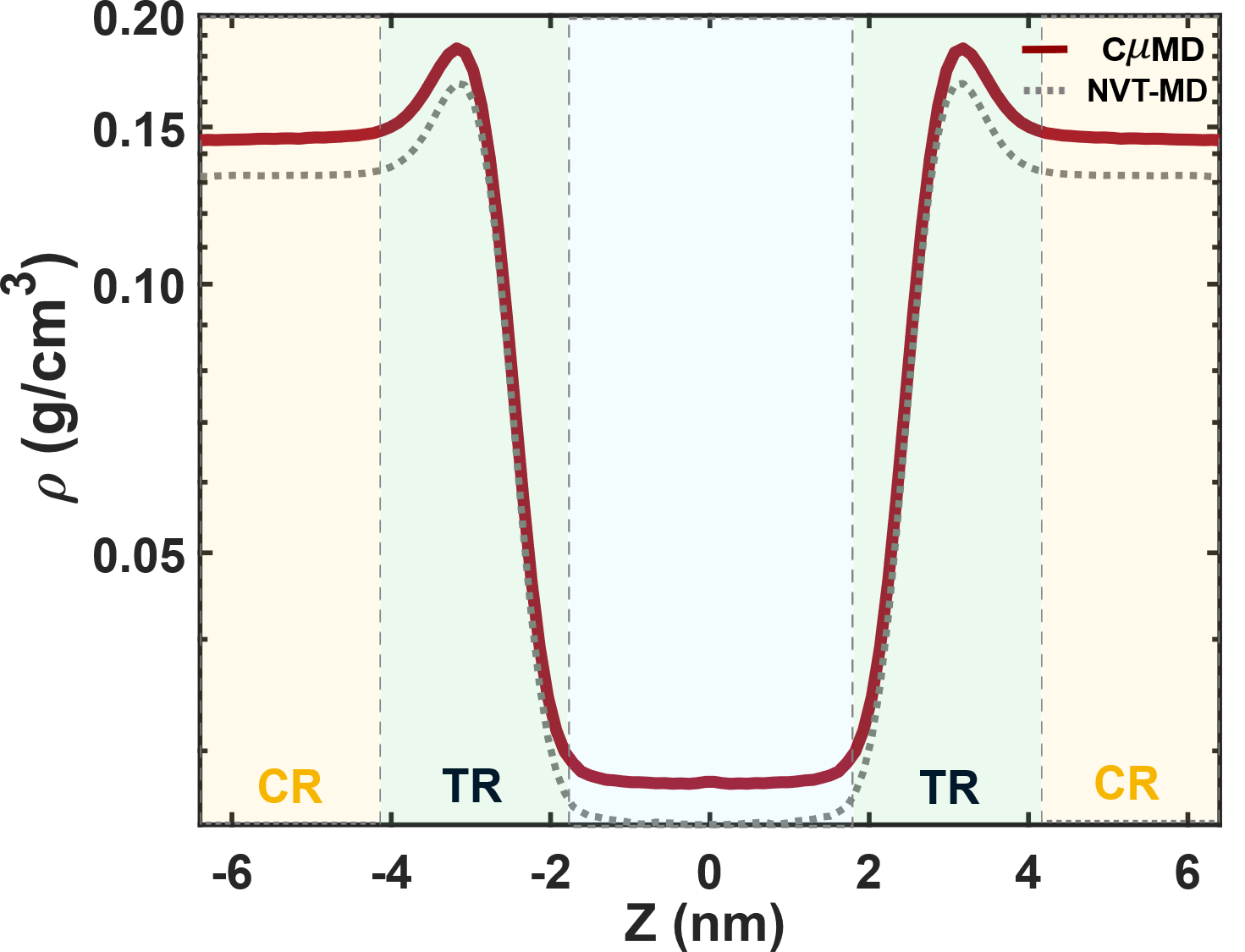}
	\centering
	\caption{Density profile of CO$_{2}$ calculated from NVT-MD and C$\mu$MD simulations at T=423 K and P=100 bar (the y-axis is in log scale).} 
	\label{density}
\end{figure}

\section{Results and Discussion}
\subsection{Gas Solubility}
The equilibrium solubility of CO$_{2}$ in water is obtained from the time averaged spatial density profile in the liquid slab. It is worth emphasizing that due to the penetration of gas molecules in the liquid slab, there is a decrement in the liquid density, which is prominent in the high pressure and negligible for the low pressure cases. 

In Fig.~\ref{density} we illustrate the density profile of CO$_{2}$ in system \emph{I} using the NVT-MD approach (dotted line). As described in the introduction, due to the solvation of CO$_{2}$ into water, the number of CO$_{2}$ in the gas phase is depleted and the vapor density is reduced by as much as $\sim$10 $\%$ and the solubility is also reduced. Our approach on the other hand keeps the vapor density at the desired value (see solid line in Fig.~\ref{density}), leading to the correct estimate. In system \emph{II} this depletion is even larger $\sim$18 $\%$ (see Fig. S2 of SI). 

We obtain for the solubility of CO$_{2}$ in system \emph{I},  m$_{CO_{2}}$=0.0140, which is in excellent agreement with the experimental data~\cite{takenouchi1964binary} (0.0135). In system \emph{II}, our calculated solubility m$_{CO_{2}}$=0.008 is in a good agreement with the theoretical result ($\sim$0.008) of Ref.~\citenum{liu2011monte}, but in disagreement with the experiment~\cite{liu2011monte}.

\subsection{Molecular Mechanism of Solvation}
In the previous section, we discussed the solubility of CO$_{2}$ and showed how using the C$\mu$MD method one can compute solubility using a relatively small system. We devote this section to a discussion on how the local interfacial surface corrugation plays a role in controlling the gas intake into the liquid. To this end, we first briefly summarize the Willard-Chandler (WC) approach~\cite{willard2010instantaneous} that we use to define the instantaneous interface. Subsequently, we investigate the hydrogen bond (HB) network of the interfacial water molecules in the vicinity of the surface. Finally, we reveal the emergence of empty patches at the interface and discuss the gas intake into the liquid. Here, we continue studying the same thermodynamic conditions as before (namely systems \emph{I} and \emph{II}).

\subsubsection{Vapor-Liquid Equilibrium Interface}

In the WC approach, the atomic density is broadened by placing a Gaussian at each atomic position. This broadened density is then collocated on a three-dimensional grid of points. The set of points from which the density is half of its bulk value density defines the interface. Here we constructed the interface by an array of 21 $\times$ 21 $\times$ 101 grid points which amounts to choosing a coarse-graining length to 2.4 \AA. For this analysis, we used the code implemented in Ref.~\citenum{sega2018pytim}.

\begin{figure}[!ht]
\includegraphics[width=0.7\textwidth]{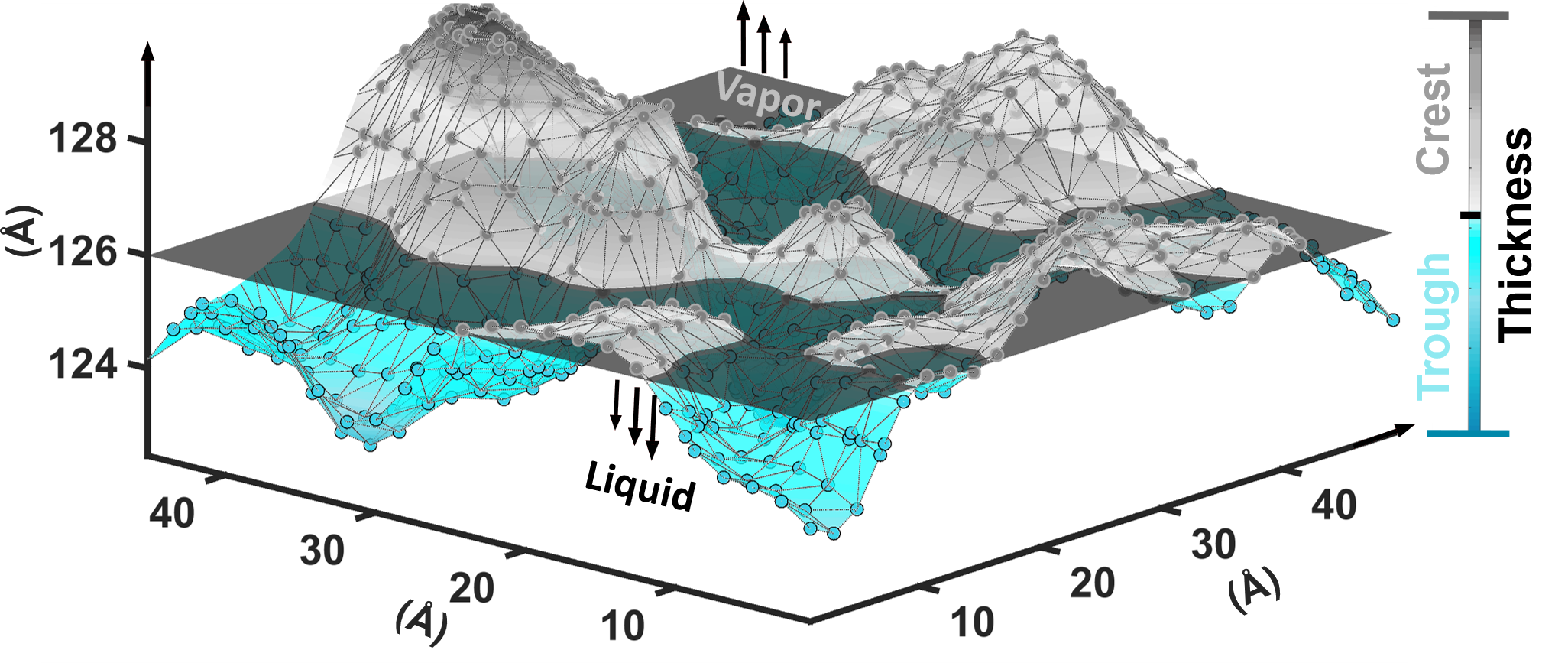}
	\centering
	\caption{The Willard-Chandler instantaneous interface. The blue and gray colors depict the trough and crest regions, and the semi-transparent gray plane shows the separation plane between these two regions. Delaunay triangles for estimation of the surface area of the interface are depicted in the wire-frame. The crest and trough thicknesses are shown on the right-side panel.} 
	\label{WC}
\end{figure}

In Fig.~\ref{WC}(a), we show one such surface. The ondulation of this surface originates from capillary wave fluctuations. We shall show below that these ondulations have a significant effect on the local water structure and the gas adsorption process. In order to carry out this analysis, we need to identify the crests and troughs of the interface. These are defined by their position relative to the instantaneous average z-coordinate (Fig.~\ref{WC}(a)).

In addition, we measure the thickness of this discretized surface as the difference in heights between the largest and smallest z-values. The interface area is obtained by first performing a Delaunay triangulation between the interface points and summing the triangles facet areas. In pure water the surface thickness increases by $\sim$50 \% in going from system \emph{II} where it is $\sim$8.5 \AA \ to system \emph{I} that has a surface thickness of $\sim$12.5 \AA. In the case of the CO$_{2}$ - H$_{2}$O system, the CO$_{2}$ molecules crowd at the interface (see Fig.~\ref{CmuMD_shem}) and increase the surface thickness by $\sim$3 \AA \ in both systems.\\

\subsubsection{HB Network in Vicinity of Surface Ondulation}

Now we investigate the influence of water surface ondulation on the interfacial water network. The instantaneous density profile of water along the z-axis, depicted in Fig. S3, shows clearly the layering of water molecules, a well-known interfacial property amply discussed in the literature~\cite{willard2010instantaneous,pezzotti20172d}.

\begin{figure}[!hb]
\includegraphics[width=0.7\textwidth]{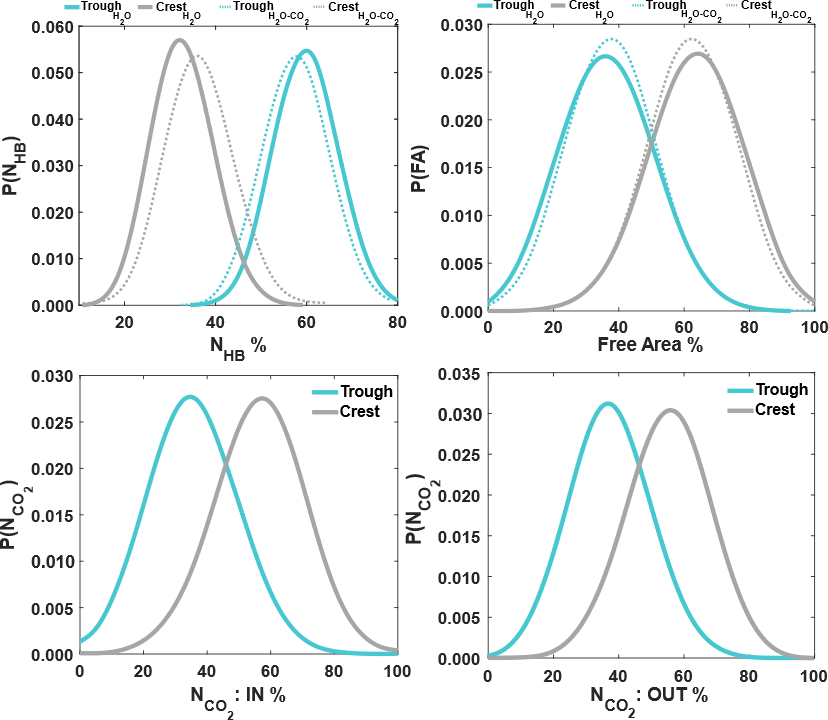}
	\centering
	\caption{(a) Percentage of hydrogen bonds that are under crest and trough regions within the first layer of water near the interface in both pure VLE and CO$_{2}$ - H$_{2}$O system, (b) the percentage of free area available under the crest and trough regions, (c) percentage of CO$_{2}$ molecules (N$_{CO_{2}}$:IN) within 2 \AA \ of crest and trough from the liquid side of the instantaneous interface, and (d) same as panel (c) from the gas side of the interface (N$_{CO_{2}}$:OUT). Here the densities of vapor CO$_{2}$ and liquid water are at T=423 K and P=100 bar.} 
\label{HB_per}
\end{figure}

Furthermore, it has been shown that the density and orientational parameters of water molecules at the first layer deviate significantly from that of the bulk~\cite{kessler2015structure}. In order to analyze this effect, we identify the surface water molecules (W$_{int}$) as those whose distance from the interfacial grid points is less than $\sim$3.5 \AA. Subsequently, the surface water molecules are classified into the crest and trough molecules using the criteria described in the SI (Section III of the SI). 

We are particularly interested in knowing the behavior of these surface water molecules and their HB network patterns across the crest and trough regions. Recently, Pezzotti et al.~\cite{pezzotti20172d} identified a  two dimensional (2D) extended HB network in this first layer of water near the interface with an average of 2.9 HB per water at 315 K and P=1 bar. We obtain an average of 2.4 and 2.7 HB per molecule for systems \emph{I} and \emph{II}, respectively. Here, we consider as hydrogen-bonded two water molecules that satisfy the geometrical criteria of O-O distance $<$ 3.5 \AA \ and H-O...O angle $<$30$^\circ$.

Fig.~\ref{HB_per}(a) illustrates that among all water molecules of the 2D HB network, the number of water molecules that participate in this network is on average higher on the trough than the crest. At T=423 K, this difference is around 30 $\%$ (see Fig.~\ref{HB}a), while at T=323 K this is around 10 $\%$ (See Fig. S5 in the SI). In other words, hydrogen bonding is more favorable on trough than in crest. This is due to the fact that the trough having a relatively higher surface area than crest can accommodate an extended HB network (See Fig S6 in the SI). This result affirms a non-uniform distribution of water molecules at the interface, and as we will discuss later, this kind of heterogeneity leads to the introduction of the surface empty patches concept.\\

\subsubsection{Surface Empty Patches and Gas Solvation Mechanism}
The concept of empty patch or cavity is well-defined in the literature and extensively used for examining the density fluctuations in bulk liquids\cite{Yagasaki2013,ohmine1999,patel2011,mittal2008,chandler2005,lum1999hydrophobicity,sosso2017,ansari2018,ansari2019,ansari2020insights}.

\begin{figure}[!hb]
\includegraphics[width=0.7\textwidth]{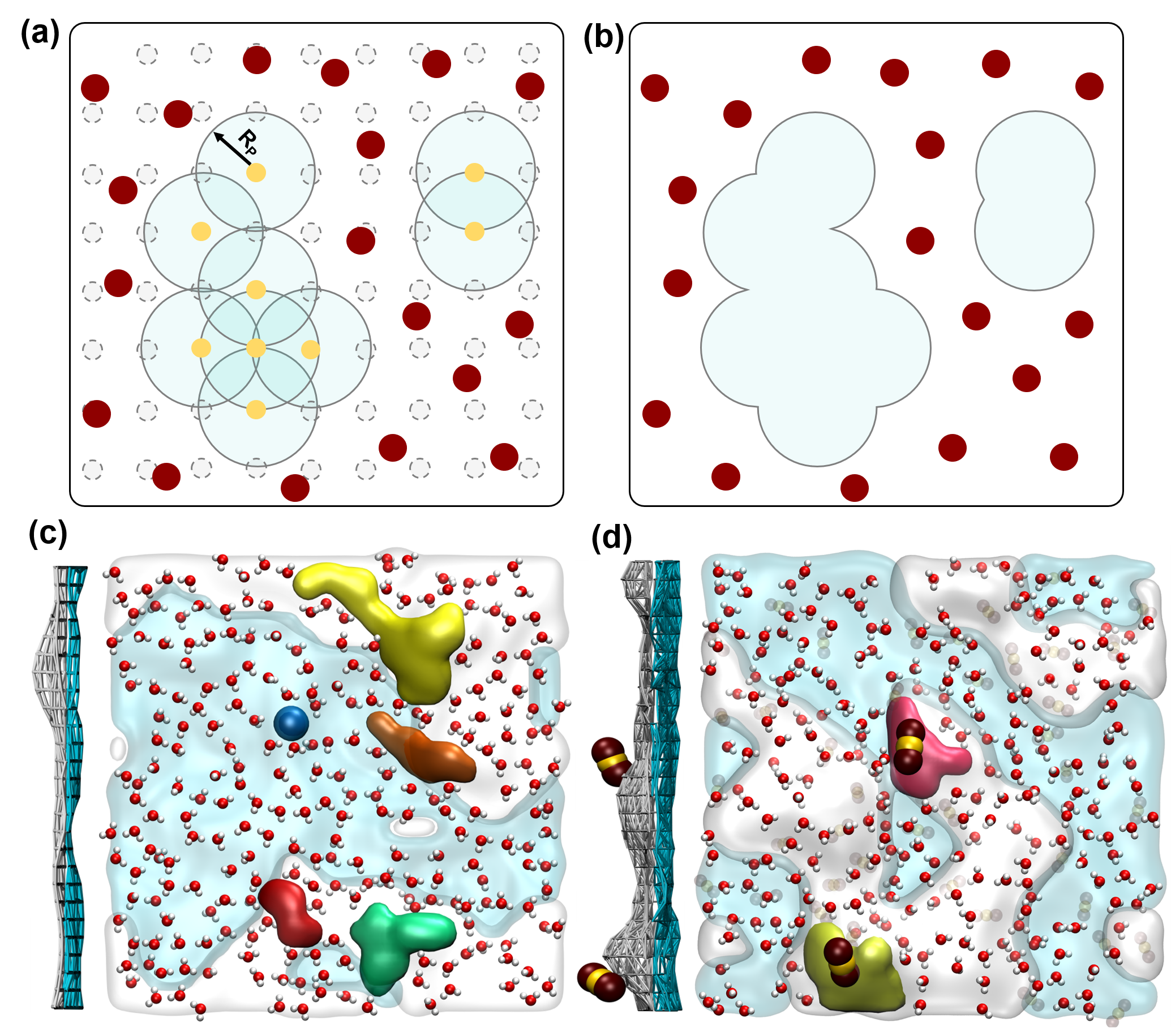}
	\centering
	\caption{ (a) Schematic representation of the empty patch identification on the surface of the WC interface. Smaller light gray circles (dotted) represent the grid points of the WC interface. Red dots indicate the oxygen atoms in the vicinity (O$_{W_{int}}$) of the grid points. The orange points are the probe centers that belong to the empty area. Semi-transparent blue circles are the probe circles to find the empty area on the surface, and (b) the empty patches are merged to indicate the total empty surface, (c) side and top views of the instantaneous interface in the pure vapor-liquid water with W$_{int}$ molecules. Empty patches are shown in solid colors, while transparent blue and gray colors depict the trough and crest regions, respectively, (d) the same as panel (c) for CO$_{2}$ - H$_{2}$O system. Two CO$_{2}$ molecules that cross the interface are shown with large red and yellow spheres. The yellow and pink empty patches on the crest region also show their entrance passage. The dissolved CO$_{2}$ molecules under the interface are shown in semi-transparent.} 
	\label{HB}
\end{figure}

In this work, for the first time, we characterize transient empty patches on the instantaneous surface.

In order to identify these patches, we draw on the interface circles of radius $R_P$ centered at the interface grid points. The union of all such circles that do not contain any water oxygen defines the surface empty patches (Fig.~\ref{WC}(b)). To do this, we use the graph algorithm implemented in Matlab\cite{MATLAB}. A similar approach was used previously in the context of void analysis in the supercooled liquid water\cite{ansari2020insights}. To define the surface empty patches area, we perform the Delaunay triangulation on a finer grid. For this analysis, we use $R_{P}$=2.5 \AA. %In the SI, we discuss how the change in R$_{P}$ can influence the results.

Fig.~\ref{HB}(a) shows some representative small and large transient empty patches on the pure vapor-liquid water interface. %Takato {\em et al.} observed a similar fragile region in the first layer of water near the hydrophobic surface~\cite{sato2018hydrophobic}.
As illustrated in Fig.~\ref{HB}(a), a major portion of the empty patch area lies on the crest. There are only a few small empty patches on the trough. We now express this observation in a quantitative way. To this purpose, we calculate the fraction of free area (sum of all empty patches area of the interface divided by the total area of the instantaneous interface) that belongs to the crest and trough. %For the vapor-liquid water interface, the percentage of the free area is around 20 $\%$ and 8.5 $\%$ of the total area of the surface at 423 K and 323 K, respectively.
Figs.~\ref{HB_per}(b) and S5(b) reveal that in the vapor-liquid water, around 70 $\%$ of the total free area is localized on the crest, and only 30 $\%$ sites on the trough.
This result is consistent with the presence of a larger number of HBs on the trough (Figs.~\ref{HB} and S5) than on the crest. Moreover, a similar percentage of the free area in CO$_{2}$ - H$_{2}$O system (see dotted lines in Fig.~\ref{HB_per} panel (b)), reveals that the presence of gas molecules near the interface has no significant effect on the organization of water molecules in the vicinity of the instantaneous interface, which is consistent with the result of Ref.~\citenum{zhang2011analysis}. Figs.~\ref{HB_per}(c) and S5(b) show that also the HB percentage on the crest and trough remains almost unaltered, especially for the lower temperature T=323 K. 

The distribution of CO$_{2}$ molecules within 2 \AA \ inside and outside of the interface (see panels (c) and (d) of Figs.~\ref{HB_per} and S5), shows that in both cases, the percentage of CO$_{2}$ molecules in the vicinity of the crest is higher than that of the trough. This suggests that the penetration of the gas molecules is more likely to take place from the crest via the interface. Panel (b) of Fig.~\ref{HB} shows two representative CO$_{2}$ molecules that penetrate the interface through the empty patches of the crest.

\section{Conclusion}

\begin{scheme}
\begin{center}
\includegraphics[width=0.7\textwidth]{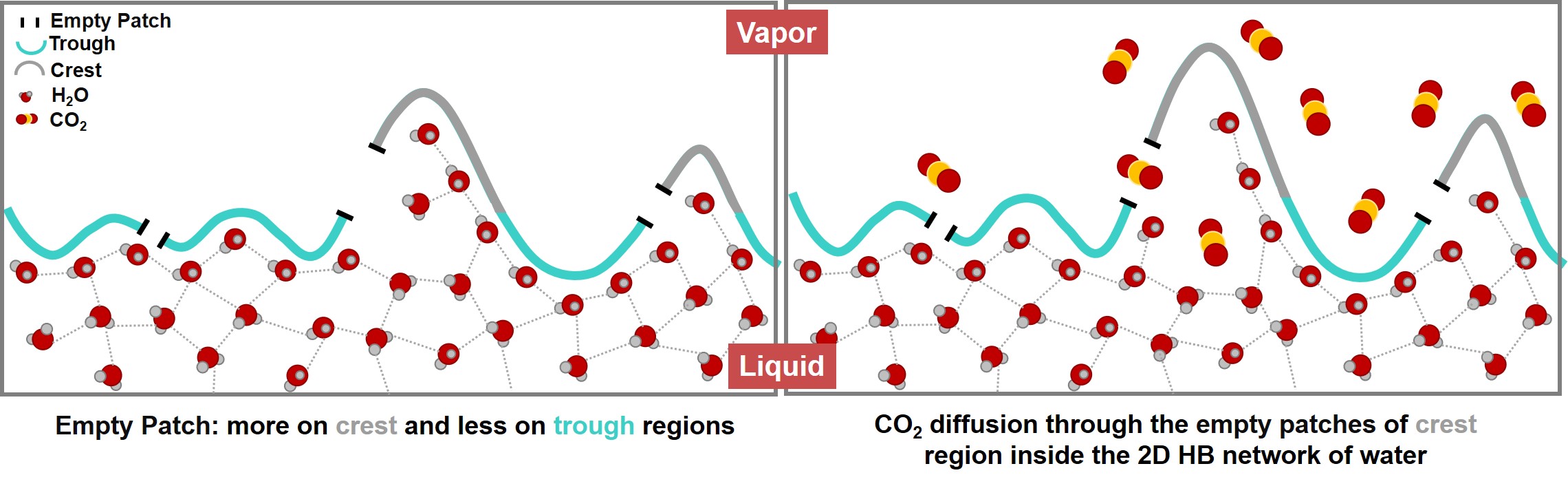}
\end{center}
\caption{Illustration of the diffusion mechanism of gas molecule through the interface.}
\label{mechanism}
\end{scheme}

In this study, we have applied the recently developed C$\mu$MD method to calculate the isothermal solubility of CO$_2$ in water at two thermodynamic conditions around the critical point of CO$_{2}$. Proper control of the gas pressure near the liquid results in an accurate calculation of this important physical property. An excellent agreement of the calculated solubility with the experimental value is obtained at temperature 423 K and pressure 100 bar. The underestimation of the solubility data at lower temperature and pressure (T=323 K and P=50 bar) is due to the force-field's inability in reproducing experimental thermodynamic conditions. More accurate potentials, such as those derived from the first principle many-body potentials or accurate machine learning-based force fields, could provide a better estimate of the calculated solubility.

Additionally, we propose the mechanism of gas molecules solvation in the aqueous medium. The schematic in Scheme I illustrates a possible diffusion process of the gas molecules through the transient empty patches of the interface. An extended 2D-HB network on the trough region of the interface makes gas molecules difficult to penetrate through them to the interface. On the contrary, the large available free area on the crest region of the interface allows a facile diffusion of the gas molecules to the bulk liquid. As far as we know, this is the first time that we quantify the diffusion mechanism of gas molecules based on the empty patches on the local curvature of the instantaneous interface. Such microscopic details are important to understand the solvation mechanism of gases, in general solutes, in different solvents. \\
\bigskip
\bigskip
\bigskip
\newpage
{\bf ORCID:}\\
Narjes Ansari: 0000-0003-2017-8431\\
Tarak Karmakar: 0000-0002-8721-6247\\
Michele Parrinello: 0000-0001-6550-3272\\

\bigskip
{\bf ACKNOWLEDGEMENTS}\\

We thank CSCS, Swiss National Supercomputing Centre for providing the computational resources. The research was supported by the European Union Grant No. ERC-2014-AdG-670227/VARMET. We also acknowledge the NCCR MARVEL, funded by the Swiss National Science Foundation.\\

\bibliographystyle{unsrt}  
\bibliography{references}{}

\end{document}